\documentclass[twocolumn,amsmath,amssymb,prb,showpacs]{revtex4}
\usepackage{epsf}

\begin{document}
\title{The inertia of stress}
\author{Rodrigo Medina}
\email{rmedina@ivic.ve}
\affiliation{Instituto Venezolano de Investigaciones Cient\'{\i}ficas, IVIC,
Apartado 21827, Caracas 1020A, Venezuela}

%\date{\today}

\begin{abstract}
We present a simple example in which
the importance of the inertial effects of stress is evident. The system is
an insulating solid narrow disc whose faces are uniformly charged with
charges of equal magnitude and opposite signs. The
motion of the system in two different directions is considered. It is shown
how the contributions to energy and momentum of the stress that develops
inside the solid to balance the electrostatic forces have to be added to
the electromagnetic contributions to obtain the results predicted
by the relativistic equivalence of mass and energy.

\vskip 10pt
{\it The following article has been accepted by the American Journal of
 Physics. After it is published, it will be found at
 http://scitation.aip.org/ajp.}
 
\end{abstract} 


\maketitle

\section{Introduction}
Recently, a proposal for the solution of the
century old problem of the self-interaction of a
charged particle was presented.\cite{Medina} One of the puzzles
of this problem\cite{Rohrlich} is that the momentum
of the electromagnetic
field of a particle with electrostatic energy $U_e$ moving with velocity
${\bf v}$ is not $U_e\gamma{\bf v}$ as required by relativity, but is
$\frac{4}{3}U_e\gamma{\bf v}$. It was shown \cite{Medina}
that the discrepancy is due to
the neglect of the inertia of the stress that is present in the particle to
balance the electrostatic repulsion. Unlike the inertia of energy, which is
well known, many physicists are not aware of the inertia of pressure
(stress). In many cases such an effect is negligible, but for the case of
the stress produced by electrostatic interactions, it is comparable
to the inertial effects of the electromagnetic fields. If the inertia of
stress is neglected, the calculations are inconsistent. In this paper we give an example in which these considerations are explicitly shown.

\section{The System}
\begin{figure}[h!]
\epsfxsize=3.3in
\centerline{\epsffile{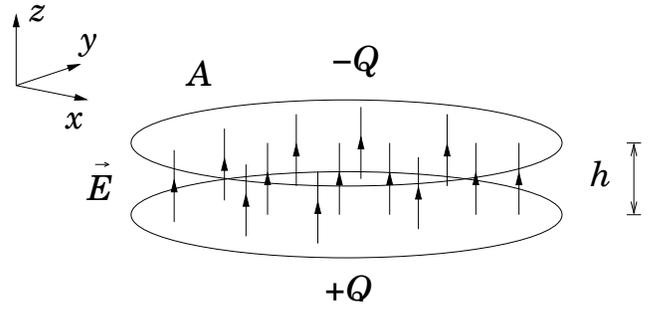}}
\caption{\label{ChargedDisc}The electric field of a charged disc.}
\end{figure}

We consider a solid disc of insulating material with
radius $R$ and thickness $h$ such that $h\ll R$ (see
Fig.~\ref{ChargedDisc}). For simplicity we assume that the material has a unit relative dielectric
constant, $\epsilon=\epsilon_0$. Both faces of the disc are uniformly charged
with opposite charges $Q$ and $-Q$.
The axis of the disc is parallel
to the $z$ axis. The lower face is positively charged.
The surface charge density of the lower face is
$\sigma=Q/A$, where $A=\pi R^2$.
If we neglect border effects, the electric field 
is ${\bf E}=(\sigma/\epsilon_0)\hat{{\bf z}}$ at any point inside the
material and zero outside. The electrostatic energy is 
\begin{equation}
U_e=\frac{\epsilon_0}{2}\!\int\! dV\,E^2=\frac{\sigma^2}{2\epsilon_0}Ah
=\frac{Q^2h}{2\epsilon_0 A}.
\end{equation}

The electrostatic interactions produce stresses in the solid disc. The stress
tensor $\tensor{\rm P}$ is defined so that the total force that the surroundings
produce on a body is the opposite of the integral of the stress
tensor over the surface of the body.\cite{StressDeff} That is,
\begin{equation}
{\bf F}= -\!\int_S\! d{\bf A}\cdot\tensor{\rm P}.
\end{equation}

\begin{figure}[h!]
\epsfxsize=3.3in
\centerline{\epsffile{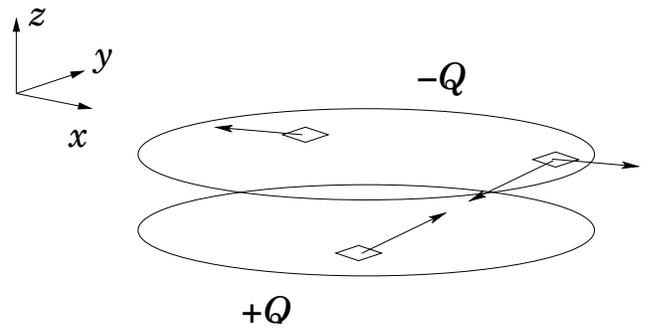}}
\caption{\label{Forces}The forces between elements of the faces. The forces are
repulsive for elements in the same face and attractive for elements in
opposite faces.}
\end{figure}

\begin{figure}[h!]
\epsfxsize=3.3in
\centerline{\epsffile{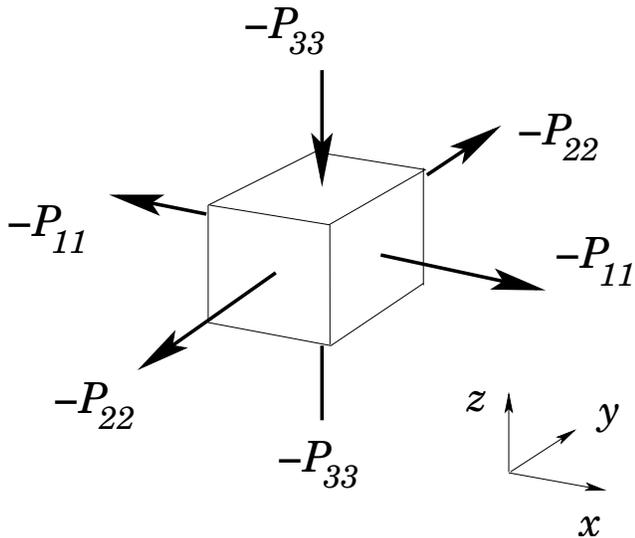}}
\caption{\label{Stresses}The stresses due to charges acting on an element of the
solid disc.}
\end{figure}

The opposite
charges in the faces attract each other, compressing the body (see
Fig.~\ref{Forces}).
A positive
stress develops for surfaces parallel to the faces. In contrast,
the repulsion between charged elements of the same face produces a radial
stretching of the body. Hence, for surfaces parallel to the $z$ axis there is
a negative stress that balances the repulsion. That is, the stress tensor
is diagonal; $P_{11}$ and $P_{22}$ are equal and negative and $P_{33}$ is
positive (see Fig.~\ref{Stresses}).

The standard way of calculating
the stress is to use Maxwell's tensor, but to keep the
exposition as elementary as possible we will determine the stress using
energy considerations. If the thickness $h$ is increased, the stress in the
$z$-direction does a work that is the opposite to the work of the compressing
force that it balances.
\begin{equation}
P_{33}A\,dh=-dW=dU_e.
\end{equation}
Then
\begin{equation}
P_{33}=\frac{1}{A}\frac{dU_e}{dh}=\frac{1}{2\epsilon_0}
\frac{Q^2}{A^2}=\frac{\sigma^2}{2\epsilon_0}.
\end{equation}

If the radius is increased, the work done by the stress is equal to the
increase in the electrostatic energy
\begin{equation}
P_{11}2\pi R h\,dR = dU_e,
\end{equation}
and
\begin{equation}
P_{11}=P_{22}=\frac{1}{2\pi Rh}\frac{dU_e}{dR}=
\frac{1}{h}\frac{dU_e}{dA}=-\frac{\sigma^2}{2\epsilon_0}.
\end{equation}
If we define $P=\sigma^2/(2\epsilon_0)$, the stress
tensor is
\begin{equation}\label{StressTensor}
\tensor{\rm P} = \begin{pmatrix}
-P&0&0\\
0&-P&0\\
0&0&P
\end{pmatrix}.
\end{equation}

Let $m_0$ be the
rest mass of the disc if it were not charged and $U_e$ be the
electrostatic energy. Then the equivalence between mass and energy predicts
that if the disc moves with velocity ${\bf v}$, its energy is 
$(m_0c^2+U_e)\gamma$ and its momentum is $(m_0+U_e/c^2)\gamma{\bf v}$,
where $\gamma=[1-(v/c)^2]^{-1/2}$.

The contributions of the electromagnetic fields to the energy and
momentum are obtained by integrating the
energy density $u$
and the Poynting vector ${\bf S}$ over the volume:
\begin{equation}
U_{\text{em}}=\!\int\! u dV=\int\!
\big(\frac{\epsilon_0}{2}E^2+\frac{1}{2\mu_0}B^2\big)dV,
\end{equation}
and
\begin{equation}
{\bf P}_{\text{em}}=\frac{1}{c^2}\!\int\! dV\,{\bf S}=
\epsilon_0\!\int\! dV\,{\bf E}\times{\bf B}.
\end{equation}

We will now evaluate the electromagnetic contributions for the disc moving
in two different directions.

\section{Motion parallel to the axis of the disk }

\begin{figure}[h!]
\epsfxsize=3.3in
\centerline{\epsffile{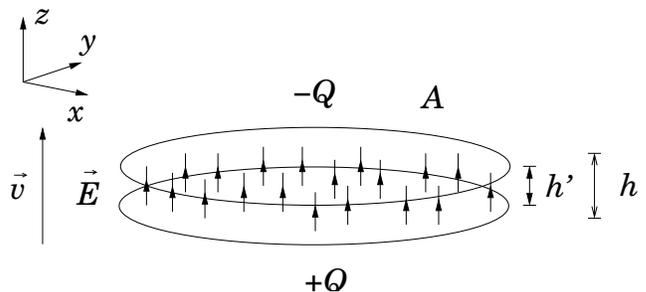}}
\caption{\label{Zmotion}Disc moving in the $z$-direction. The thickness is
reduced by the Lorentz contraction, $h^{\prime}=h\gamma^{-1}$. There is no
magnetic field and therefore there
is no electromagnetic contribution to the momentum. The electric field
is the same, but the energy is reduced by the volume contraction.}
\end{figure}

Consider the disc moving with velocity ${\bf v}=v\hat{\bf z}$ (see
Fig.~\ref{Zmotion}).
Quantities for the moving body are denoted by a prime.
The circular faces remain the same, $A^\prime=A$, so the charge density is
the same $\sigma^\prime=\sigma$, but the thickness is reduced by the
Lorentz contraction $h^\prime=h\gamma^{-1}$. The electric field is 
also the same ${\bf E}^\prime={\bf E}$.
Inside the disc $ \partial {\bf E}/\partial t=0$, so there is no
magnetic field ${\bf B}^\prime=0$. Then $u^\prime=u$ and ${\bf S}^\prime=0$.
Finally
\begin{align}
U_{\text{em}}&=u^\prime Ah^\prime=uAh\gamma^{-1}=U_e\gamma^{-1}, \\
\noalign{\noindent and}
{\bf P}_{\text{em}}&=0.
\end{align}

We see that there is no contribution to the momentum even though we might have
expected
$c^{-2}U_e\gamma{\bf v}$ and that the energy decreases as $\gamma^{-1}$
instead of increasing as $\gamma$. Considering only the
electromagnetic fields does not give the correct result.

\section{Motion perpendicular to the axis of the disk }
\begin{figure}[h!]
\epsfxsize=3.3in
\centerline{\epsffile{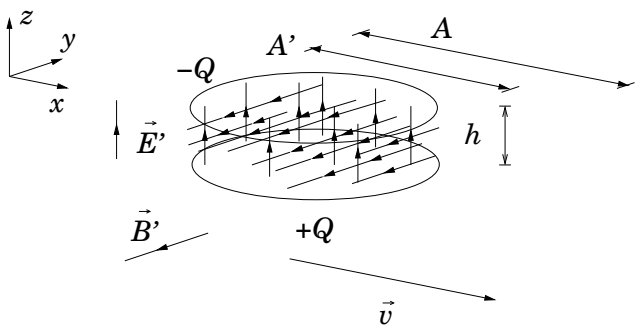}}
\caption{\label{Xmotion}Disc moving in the $x$-direction. The thickness is
the same, but the area of the faces is reduced by Lorentz contraction. The
electric field is increased because of the increase in the surface charge
density, $\sigma^{\prime}=\sigma\gamma$. There is a magnetic field
${\bf B}^{\prime}=c^{-2}{\bf v}\times{\bf E}^{\prime}$. Therefore
there is a contribution to the momentum.}
\end{figure}

Now consider the disc moving with velocity ${\bf v}=v\hat{\bf x}$
(see Fig.~\ref{Xmotion}).
The thickness is the same $h^\prime=h$, but the faces become elliptical
because of Lorentz contraction in the $x$-direction. The area
is reduced $A^\prime=A\gamma^{-1}$. The charge density is increased
$\sigma^\prime=\sigma\gamma$ and so is the electric field ${\bf E}^\prime=
\gamma{\bf E}$. There is also a magnetic field inside the disc
produced by the two sheets of opposite currents. The field can be calculated
using Amp\`ere's law
\begin{equation}
{\bf B}^\prime=-\mu_0\sigma\gamma v\hat{\bf y}=
\frac{1}{c^2}{\bf v}\times{\bf E}^{\prime}.
\end{equation}

The Poynting vector is
\begin{equation}
{\bf S}^\prime=\frac{1}{\mu_0}{\bf E}^\prime\times{\bf B}^\prime=
\frac{1}{\epsilon_0}(\sigma\gamma)^2{\bf v},
\end{equation}
and the electromagnetic momentum is
\begin{equation}
{\bf P}_{\text{em}}=\frac{Ah\gamma^{-1}}{c^2}{\bf S}^\prime=
2\frac{U_e}{c^2}\gamma{\bf v}.
\end{equation}

The electromagnetic energy is
\begin{subequations}
\begin{align}
U_{\text{em}}&= u^\prime
Ah\gamma^{-1}=\big[\frac{1}{2\epsilon_0}(\sigma\gamma)^2+
\frac{\mu_0}{2}(\sigma\gamma v)^2\big]Ah\gamma^{-1} \\
&= U_e\gamma\Big[1+\big(\frac{v}{c}\big)^2\Big].
\end{align}
\end{subequations}

In this case the result is also not as might be expected. The energy has an extra
$(v/c)^2$ term and the momentum is twice the expected value. 
The energy and momentum of the electromagnetic field do not form a
four-vector, and the effective mass is anisotropic.
Something is missing and that is the inertia of stress.

\section{The Inertia of Stress}
We will use the following relativistic conventions,
$x^0=-x_0=ct$, $x^1=x_1=x$, $x^2=x_2=y$, and $x^3=x_3=z$. Greek
indices take the values 0--3,
and Latin indices take the values 1--3. The unit vectors in the direction
of the spatial axes are $\hat{\bf e}_i$.
Repeated indices indicate an implicit sum.
The four-velocity $u^\alpha$ is
related to the velocity ${\bf v}=v_i\hat{\bf e}_i$ by, $u^i=\gamma v_i$ and
$u^0=c\gamma$.

The relativistic dynamics of a continuous
medium is ruled by the energy and momentum conservation
equation\cite{EqRelativ}
\begin{equation}\label{RelEq}
\nabla_{\alpha}(\Theta^{\alpha \beta}+ P^{\alpha \beta})=f^{\beta},
\end{equation}
where $f^{\beta}$ is the force density four-vector, $\Theta^{\alpha \beta}$
is the energy, and momentum density four-tensor, and $P^{\alpha \beta}$ is the
stress four-tensor. Both $\Theta^{\alpha \beta}$ and $P^{\alpha \beta}$ are
symmetric tensors. The spatial components of $f^{\beta}$ form the force density,
${\bf f}=f^i\hat{\bf e}_i$.
The temporal component is proportional to the power density,
$f^0=({\bf f}\cdot{\bf v})/c$. The energy-momentum tensor
is obtained from the four-velocity by
\begin{equation} 
\Theta^{\alpha \beta}=\tilde\mu u^{\alpha}u^{\beta}.
\end{equation} 
The four-scalar $\tilde\mu$ is the density of the rest mass
with respect to the proper
volume (the volume of an element at rest). The usual rest mass density
is $\mu=\tilde\mu\gamma$. The spatial part of $\Theta^{\alpha \beta}$ is
the momentum current density $\Theta^{ij}=\mu\gamma v_i v_j$, and $\Theta^{00}$
is the energy density, $\Theta^{00}=\mu c^2\gamma$. The other elements of
$\Theta^{\alpha \beta}$ with only one temporal index are the energy
current density vector
$c\,\Theta^{i0}\hat{\bf e}_i$ and the momentum density vector
$c^{-1}\Theta^{0i}\hat{\bf e}_i$.

The stress four-tensor $P^{\alpha \beta}$ reduces to the purely spatial
stress tensor when the matter element is at rest. When the element is
moving, there are temporal components that contribute to the energy density
$P^{00}$ and to the momentum density $P^{0i}/c$. That is, the stress has
inertial effects. Because $u^{\alpha}$ is purely temporal at rest, we have
\begin{equation}\label{condition}
P^{\alpha \beta}u_{\beta}=0.
\end{equation}
Equation (\ref{condition}) is valid in any reference frame and can be
used to obtain the
temporal components of the stress, which are
\begin{subequations}
\begin{align}
P^{i0}&=\frac{1}{c}P^{ij}v_j ,\\
\noalign{\noindent and}
P^{00}&=\frac{1}{c^2}P^{ij}v_i v_j,
\end{align}
\end{subequations}
where $P^{ij}$ is the stress tensor in that frame.

By separating the spatial and temporal components, Eq.~(\ref{RelEq})
reduces to the momentum and power equations,
\begin{equation}
\label{momentum}
\frac{\partial}{\partial t}(\mu\gamma{\bf v} + \tensor{\rm P}\cdot{\bf
v}/c^2)+
\nabla\cdot(\mu\gamma{\bf v}{\bf v} +\tensor{\rm P})={\bf f},
\end{equation}
and
\begin{equation}
\label{power}
\frac{\partial}{\partial t}(\mu c^2\gamma +
{\bf v}\cdot\tensor{\rm P}\cdot{\bf v}/c^2)+
\nabla\cdot(\mu c^2\gamma{\bf v} +\tensor{\rm P}\cdot{\bf v})=
{\bf f}\cdot{\bf v}.
\end{equation}

It is interesting to compare these equations with the non-relativistic ones,
which are\cite{EqClass}
\begin{equation}
\frac{\partial}{\partial t}(\mu{\bf v})+
\nabla\cdot(\mu{\bf v}{\bf v} +\tensor{\rm P})={\bf f},
\end{equation}
and
\begin{equation}
\frac{\partial u}{\partial t}+
\nabla\cdot(u{\bf v}+\tensor{\rm P}\cdot{\bf v})=
{\bf f}\cdot{\bf v}.
\end{equation}
Here $u$ is the energy density of matter which includes the kinetic and
internal energies. Note that the only contributions of the stress which
appear in the non-relativistic equations are those in the divergence 
terms. Also note that the stress that multiplies the velocity in the
time derivative of
Eq.~(\ref{momentum}) does not vanish in the small velocity
limit ($v/c\to 0$). This case is an example in which the non-relativistic
limit ($c\to\infty$) is different from the small velocity limit.
That is, the inertia of stress is a purely relativistic
phenomenon, which does not have a Newtonian explanation.

\section{Stress contributions to energy and momentum}

We now calculate the contributions of stress to the energy and momentum.
The contributions are
\begin{align}
U_S&=\frac{1}{c^2}\!\int\! dV\,{\bf v}\cdot\tensor{\rm P}\cdot{\bf v}\\
 & = \!\int\! dV\,P^{00}, \\
\noalign{\noindent and}
{\bf P}_S&=\frac{1}{c^2}\!\int\! dV\,\tensor{\rm P}\cdot{\bf v}\\
 & = \frac{1}{c}\!\int\! dV\,(P^{0i}\hat{\bf e}_i).
\end{align}

The easiest way to obtain the stress tensor for the moving disc is to
use the Lorentz transformation of the result for the rest frame,
Eq.~(\ref{StressTensor}):
\begin{equation}
P^{\prime\mu\nu}=L^{\mu}_{\alpha}L^{\nu}_{\beta}P^{\alpha\beta},
\end{equation}
where for motion in the $z$-direction, the transformation
matrix is
\begin{equation}
(L^{\mu}_{\nu})= \begin{pmatrix}
\gamma&0&0&\beta\gamma\\
0&1&0&0\\
0&0&1&0\\
\beta\gamma&0&0&\gamma
\end{pmatrix},
\end{equation}
where $\beta=v/c$, and the four-tensor of stress is
\begin{equation}
(P^{\prime\mu\nu})=\begin{pmatrix}
\beta^2\gamma^2P&0&0&\beta\gamma^2P\\
0&-P&0&0\\
0&0&-P&0\\
\beta\gamma^2P&0&0&\gamma^2P
\end{pmatrix}.
\end{equation}
The energy of stress is 
\begin{equation}
U_S= (\beta\gamma)^2PAh^{\prime}=\big(\frac{v}{c}\big)^2\gamma U_e.
\end{equation}
The total energy is then
\begin{subequations}
\begin{align}
U&=U_S+U_{\text{em}}=\big(\frac{v}{c}\big)^2\gamma U_e + \gamma^{-1}U_e\\
&= \Big[\big(\frac{v}{c}\big)^2+\gamma^{-2}\Big]\gamma U_e=\gamma U_e,
\end{align}
\end{subequations}
as expected.

The momentum of stress is
\begin{equation}
{\bf P}_S=\frac{1}{c}\beta\gamma^2PAh^{\prime}\hat{\bf z}=
\frac{U_e}{c^2}\gamma{\bf v}.
\end{equation}
The total momentum is also the expected value, 
${\bf P}={\bf P}_S+{\bf P}_{\text{em}}=(U_e/c^2)\gamma{\bf v}$.

Now let us consider motion in the $x$-direction. The matrix
of the Lorentz transformation is
\begin{equation}
(L^{\mu}_{\nu})=\begin{pmatrix}
\gamma&\beta\gamma&0&0\\
\beta\gamma&\gamma&0&0\\
0&0&1&0\\
0&0&0&1
\end{pmatrix},
\end{equation}
and the stress is
\begin{equation}
(P^{\prime\mu\nu})= \begin{pmatrix}
-\beta^2\gamma^2P&-\beta\gamma^2P&0&0\\
-\beta\gamma^2P&-\gamma^2P&0&0\\
0&0&-P&0\\
0&0&0&P\\
\end{pmatrix}
.
\end{equation}
The energy of stress is
\begin{equation}
U_S= -(\beta\gamma)^2PA^{\prime}h=-\big(\frac{v}{c}\big)^2\gamma U_e.
\end{equation}
The total energy is therefore
\begin{equation}
U=U_{\text{em}}+U_S=\gamma U_e\Big[1+\big(\frac{v}{c}\big)^2\Big]
-\big(\frac{v}{c}\big)^2\gamma U_e=\gamma U_e,
\end{equation}
which is the expected result.

The momentum of stress is
\begin{equation}
{\bf P}_S=-\frac{1}{c}\beta\gamma^2PA^{\prime}h\hat{\bf x}=
-\frac{U_e}{c^2}\gamma{\bf v}.
\end{equation}
In this case the correct total momentum is
\begin{equation}
{\bf P}={\bf P}_{\text{em}}+{\bf P}_S= \frac{2U_e}{c^2}\gamma{\bf v}-
\frac{U_e}{c^2}\gamma{\bf v}=\frac{U_e}{c^2}\gamma{\bf v}.
\end{equation}
Everything works if we take into account the contributions of
stress.

The energy and momentum of an extended system are obtained by integrating
the energy density and the momentum density over the entire volume.
The spaces of two different reference frames are different three-dimensional
hyperplanes in the four-dimensional Minkowski space. So, in principle,
the total energy and momentum of an extended system in different frames
are not the same physical quantities.
If energy and momentum are conserved, they form a four-vector.
When the electromagnetic field is not free, that is, when there
are charges and currents, the field itself is stressed. This stress
contributes to the energy and momentum of the field, which is why
these quantities do not form a four-vector. The electromagnetic forces
produce stresses in matter. The stress of matter also contributes
to the energy and momentum, and also do not form a
four-vector. The contributions of the stress of matter are exactly
opposite to those of the stress of the field. Thus, if the contributions
of the stress of matter are added to those of the field, the total energy
and momentum transform as a four-vector.

\section*{Acknowledgments}
I wish to thank Dr.~Victor Villalba and Dr.~Enrnesto Medina for many
very useful discussions and for reading the manuscript.

\end{document}